\documentclass{icrc29}
\usepackage{graphicx,amssymb,amsmath,times}
\usepackage{wrapfig}
\begin{document}
\title[Effects of the tilted and wavy HCS on the solar modulation of
GCRs]{Effects of the tilted and wavy current sheet on the solar
modulation of galactic cosmic rays}
\author[Shoko Miyake and Shohei Yanagita] {Shoko Miyake and Shohei Yanagita\\
        Faculty of Science, Ibaraki University, Mito 310-8512, Japan
        }
\presenter{Presenter: Shoko Miyake (miyakesk@cc.nao.ac.jp), \  
jap-miyake-S-abs1-sh34-oral}

\maketitle

\begin{abstract}

Transport equation of the galactic cosmic ray (GCR) is numerically
 solved for $qA>0$ and $qA<0$ based on the stochastic differential
 equation (SDE) method. We have developed a fully time-dependent and
 three-dimensional code adapted for the wavy heliospheric current sheet
 (HCS). Results anticipated by the drift pattern are obtained for sample
 trajectories and distributions of arrival points at the heliospheric
 boundary for GCR protons. Our simulation reproduced a 22-year cycle of
 solar modulation which is qualitatively consistent with
 observations. Energy spectra of protons at 1 AU are calculated and
 compared with the observation by BESS.

\end{abstract}

\section{Introduction}
 
Transport of the galactic cosmic rays in the heliosphere is described by
Parker's transport equation. The Parker's equation is equivalent to a
coupled stochastic differential equation (SDE) [1, 2]. This SDE method
allows us to get some information about solar modulation phenomena of
galactic cosmic ray (GCR) not obtained by other numerical methods, such
as distributions of arrival time, energy lost, and trajectory. We have
developed a fully time-dependent and three-dimensional code based on the
SDE method. We present some numerical results on solar modulation
effects on GCR obtained by this new code. 
 
\section{The Model}
 
\begin{figure}[t]
  \begin{minipage}{0.3\hsize}
  \begin{center}
    \includegraphics[bb = 0mm 0mm 200mm 60mm, scale  =0.28]{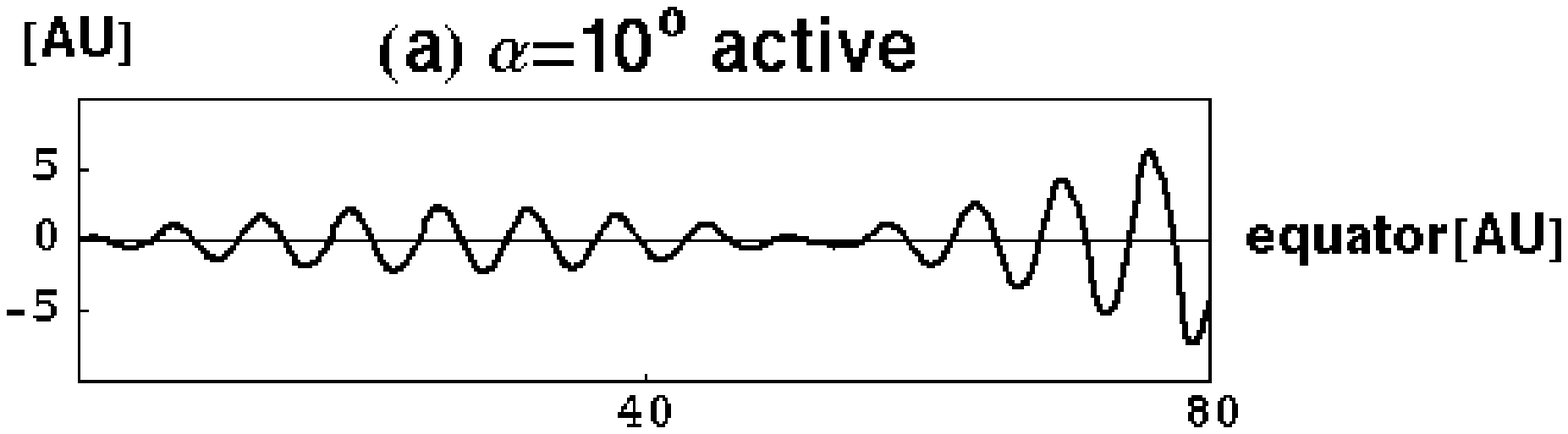}
    \\
    \includegraphics[bb = 0mm 0mm 200mm 145mm, scale  =0.28]{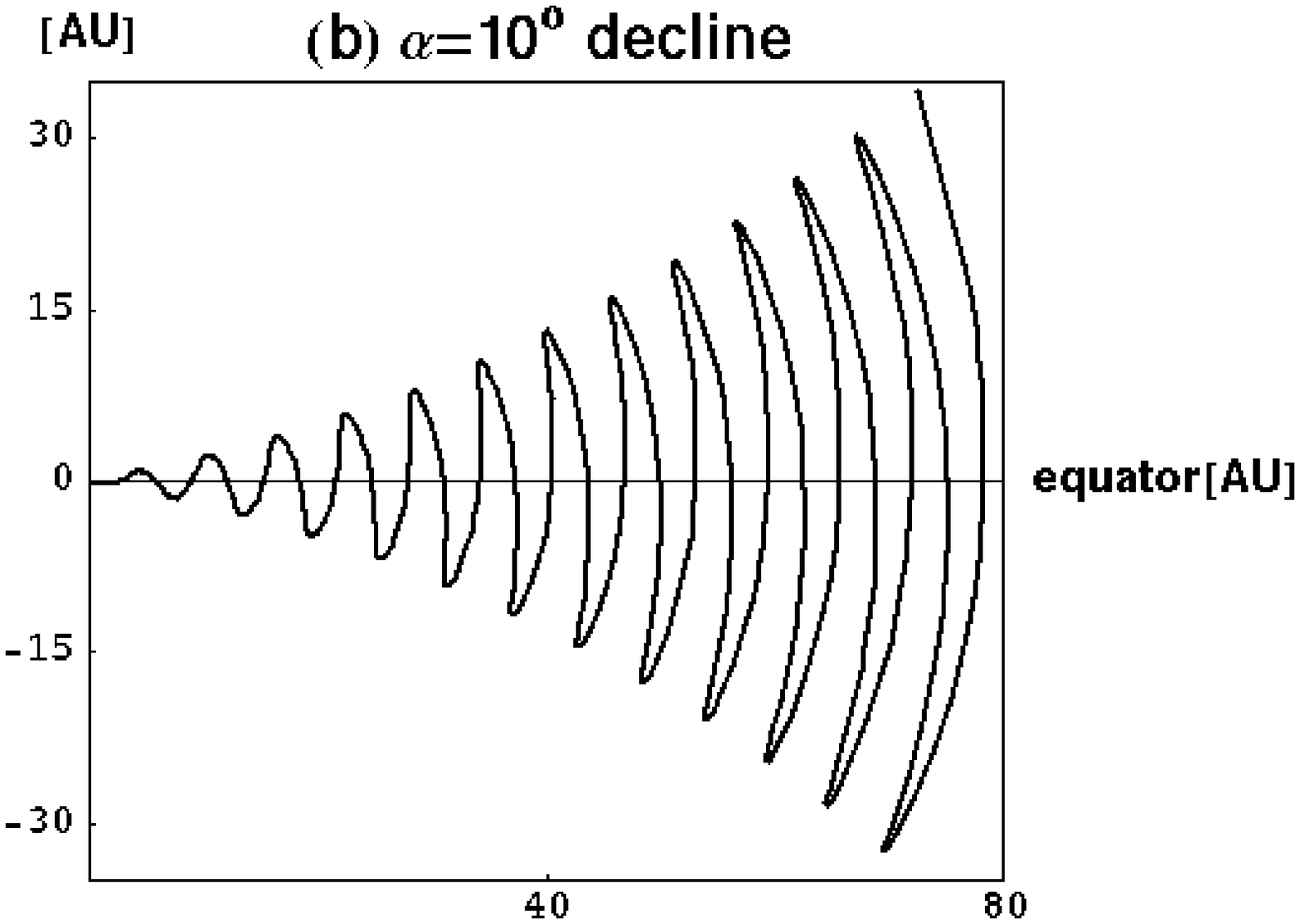}
  \end{center}
  \end{minipage}
  \begin{minipage}{0.4\hsize}
  \begin{center}
    \includegraphics[bb = 0mm 0mm 200mm 275mm, scale =  0.21]{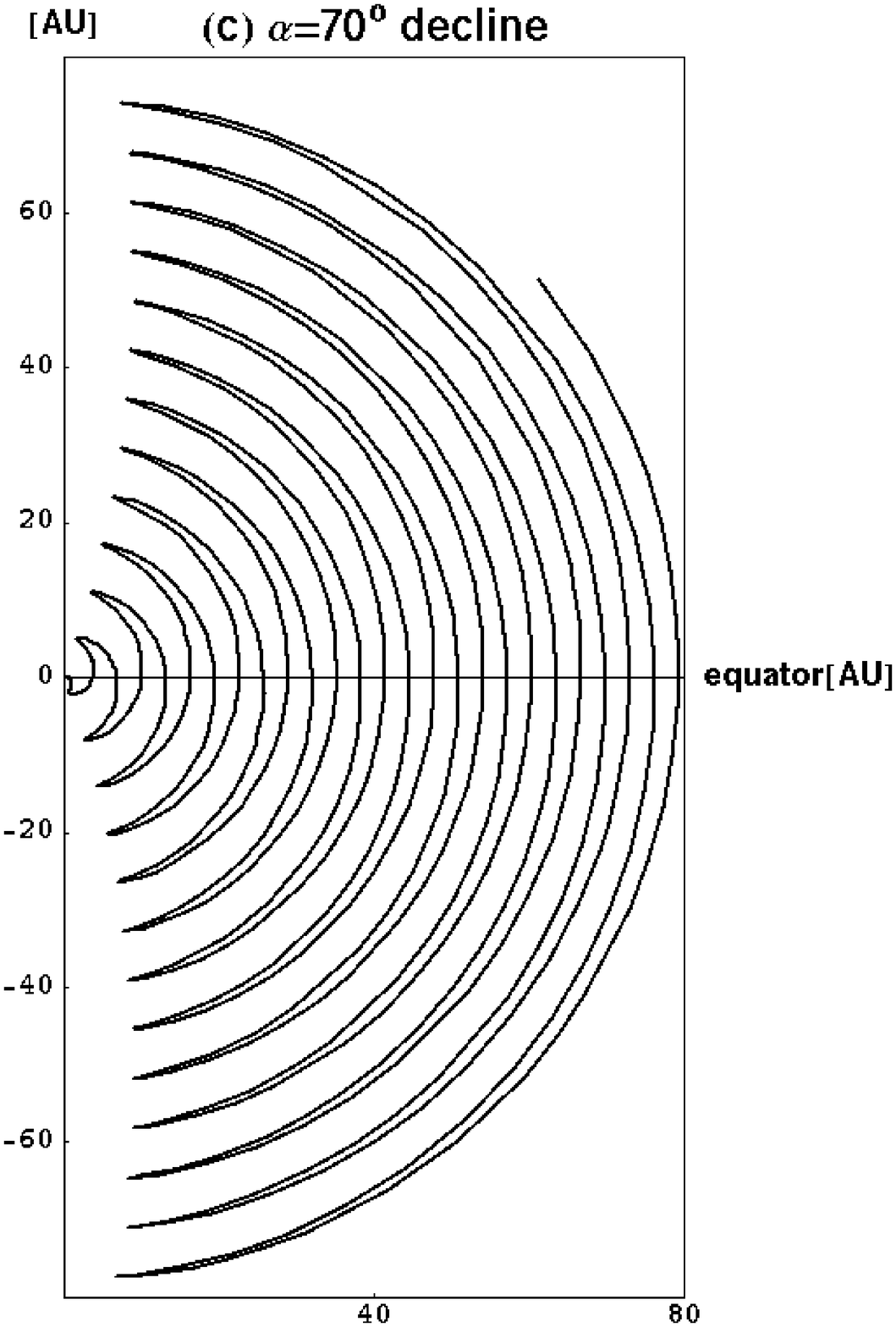}
  \end{center}
  \end{minipage}
  \begin{minipage}{0.25\hsize}
  \caption{(a) A view of the HCS projected onto a meridional plane as a
   function of radial distance from the Sun. The tilt angle near the Sun
   is $10^\mathrm{o}$, and the solar activity heading for the maximum;
   (b)   Same as (a), but the solar activity heading for the minimum;(c)
   Same as (b), but the tilt angle near the Sun is $70^\mathrm{o}$.}
 \label{cs}
 \end{minipage}
\end{figure}

The SDE equivalent to Parker's transport equation is written as \cite{Zhang} 
\begin{equation}
 \begin{array}{llllll}
d{\bf{X}} & = & \left( \boldsymbol{\nabla} \cdot {\boldsymbol{\kappa}} 
  - {\bf{V}} - {\bf{V_d}} \right)dt + \sum_{s} {\boldsymbol{\sigma}}_{s} 
  dW_{s}(t)~, \\
   & &\\
dP & = & \frac{1}{3} P\left( \boldsymbol{\nabla} \cdot {\bf{V}} \right)dt~,
  \end{array}
\label{sde}
\end{equation}
where $\bf{X}$ and $P$ are the position and the momentum of the
particle, $\bf{V}$ is the solar wind velocity, $\bf{V_d}$ is the
gradient-curvature drift velocity, ${\boldsymbol{\kappa}}$ is the
diffusion coefficient tensor, $\sum_{s} {\sigma}^{\mu}_{s}
{\sigma}^{\nu}_{s} = 2 {\kappa}^{\mu \nu}$, and $dW_{s}$ is a Wiener
process given by the Gaussian distribution. We adopted $V = 400$ km/s,
${\kappa}_{\parallel}=1.5 \times 10^{21} \beta (p/(1GeV/c))(Be/B)$
$\mathrm{cm^2/s}$, ${\kappa}_{\bot}=0.05 {\kappa}_{\parallel}$, $B$ is
the magnetic field, and $Be = 5 \mathrm{nT}$ is $B$ at the earth. We
assume the structure of heliospheric magnetic field (HMF) as standard
Parker spiral. We have used the ``drift velocity field method''
\cite{Drift} for the calculation of drift in the heliospheric current
sheet (HCS). In our simulation, particles start at 1 AU on the
equatorial plane and run backward in time until they exit the
heliospheric boundary, 80 AU.

We assume the structure of the HCS is as
\begin{equation}
\theta = \dfrac{\pi}{2} - \mathrm{sin}^{-1}\left[ \left|
    \mathrm{sin} \left( \dfrac{\pi}{2} +
    \pi\dfrac{t-\frac{r-r_{\odot}}{V}}{11 \mathrm{years}} \right) 
    \right|~\mathrm{sin}\left( \phi - {\phi}_0 - {\Omega}_{\odot}t
    + \dfrac{\left( r - r_{\odot} \right) {\Omega}_{\odot}}{V} \right) 
    \right]~,
\label{theta}
\end{equation}
where r, $\theta$, $\phi$ are spherical polar coordinates relative to the
solar rotation axis, ${\phi}_0$ is a constant, $r_{\odot}$ is the radius
of the Sun, and ${\Omega}_{\odot}$ is the rotational angular velocity of
the Sun. Similar expression for HCS had been adopted by Jokipii and Thomas
\cite{Sheet}, but with constant tilt angles contrary to our
model. In this simple HCS model, the tilt angle changes
continuously from $90^\mathrm{o}$ to $0^\mathrm{o}$ to $90^\mathrm{o}$
over a period of 11 year. The solar polarity of the HMF reverses when
the tilt angle of HCS near the sun reaches $90^\mathrm{o}$. Figure
\ref{cs} shows views of HCS projected onto a meridional plane for three
phases in the solar cycle. We have not taken into account in our model
such phenomena as the global merged interaction regions, and the
dynamically changing HCS with Fisk-like field \cite{Kota}.

\begin{wrapfigure}[15]{r}{60mm}
 \begin{center}
  \vspace{-20mm}
  \includegraphics[bb = 0mm 10mm 230mm 200mm, scale  =0.27]{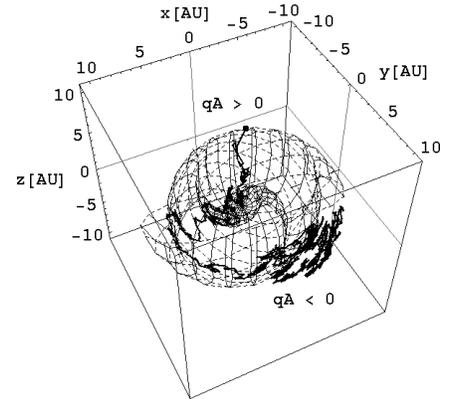}
 \caption{Sample trajectory for two protons. The tilt angle
   near the Sun is $40^\mathrm{o}$, and the solar activity heading for
  the maximum. The energy at 1 AU on the equatorial plane is 1 GeV.}
 \label{trajectory}
 \end{center}
\end{wrapfigure}

\section{Results}

Figure \ref{trajectory} shows a sample trajectory for two protons which
propagate from a point at 1 AU on the equatorial plane to 10 AU. The
coordinate system of the figure is corotating with the sun. The drift
pattern of the GCR depends on the polarity $qA$, where $q$ is a charge
sign of the sample particle and $A$ is a constant characterizing
HMF. For the positive polarity $qA>0$, the particles drift from the
polar region towards the HCS, and outward along the HCS. In contrast,
for the negative polarity $qA<0$, the direction of the drift is opposite
to the case for $qA>0$.

\begin{figure}[t]
  \begin{minipage}{0.5\hsize}
  \begin{center}
 \includegraphics[bb = 0mm 45mm 200mm 230mm, scale = 0.36]{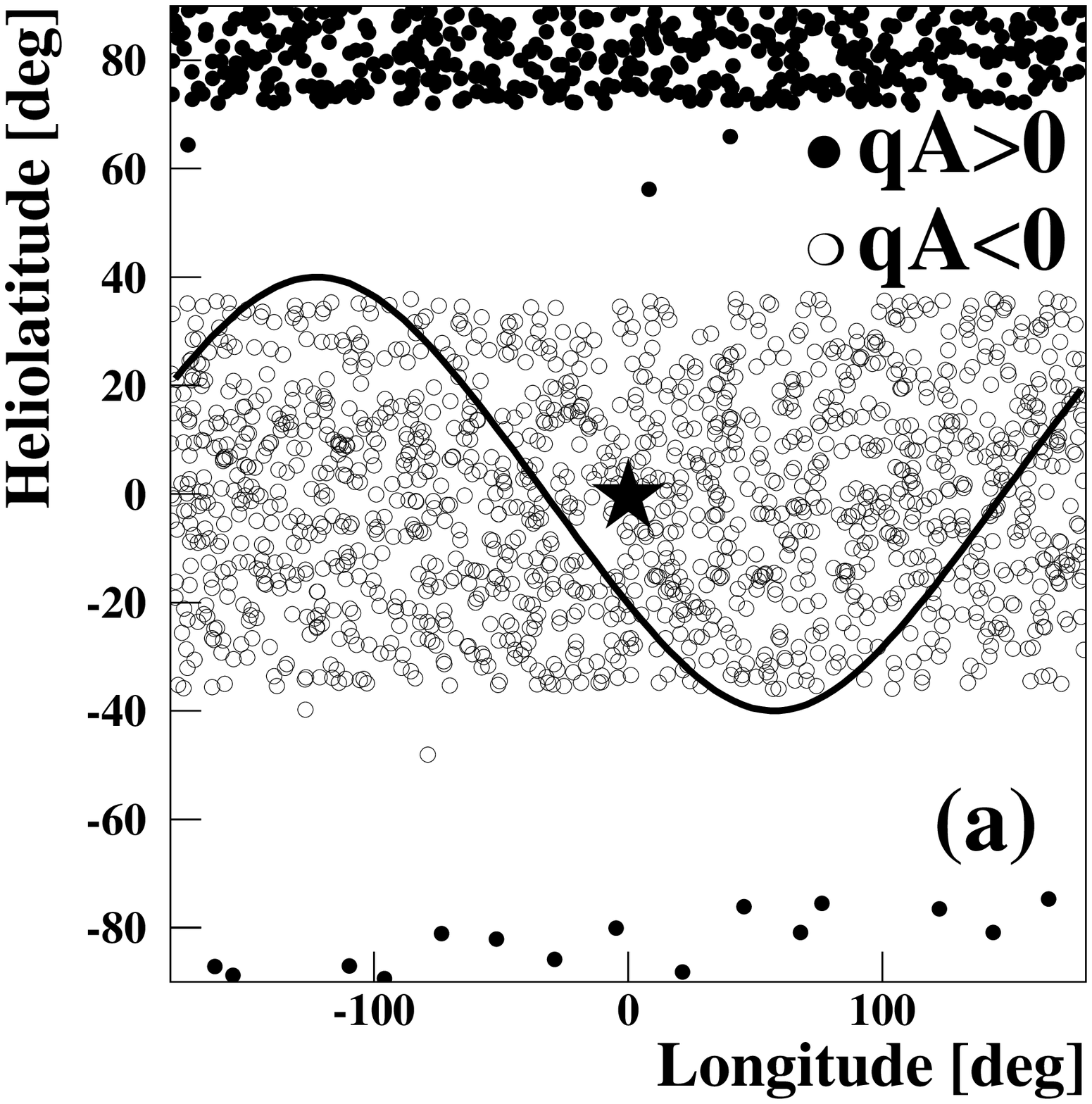}
  \end{center}
  \end{minipage}
  \begin{minipage}{0.5\hsize}
  \begin{center}
 \includegraphics[bb = 0mm 45mm 200mm 230mm, scale = 0.36]{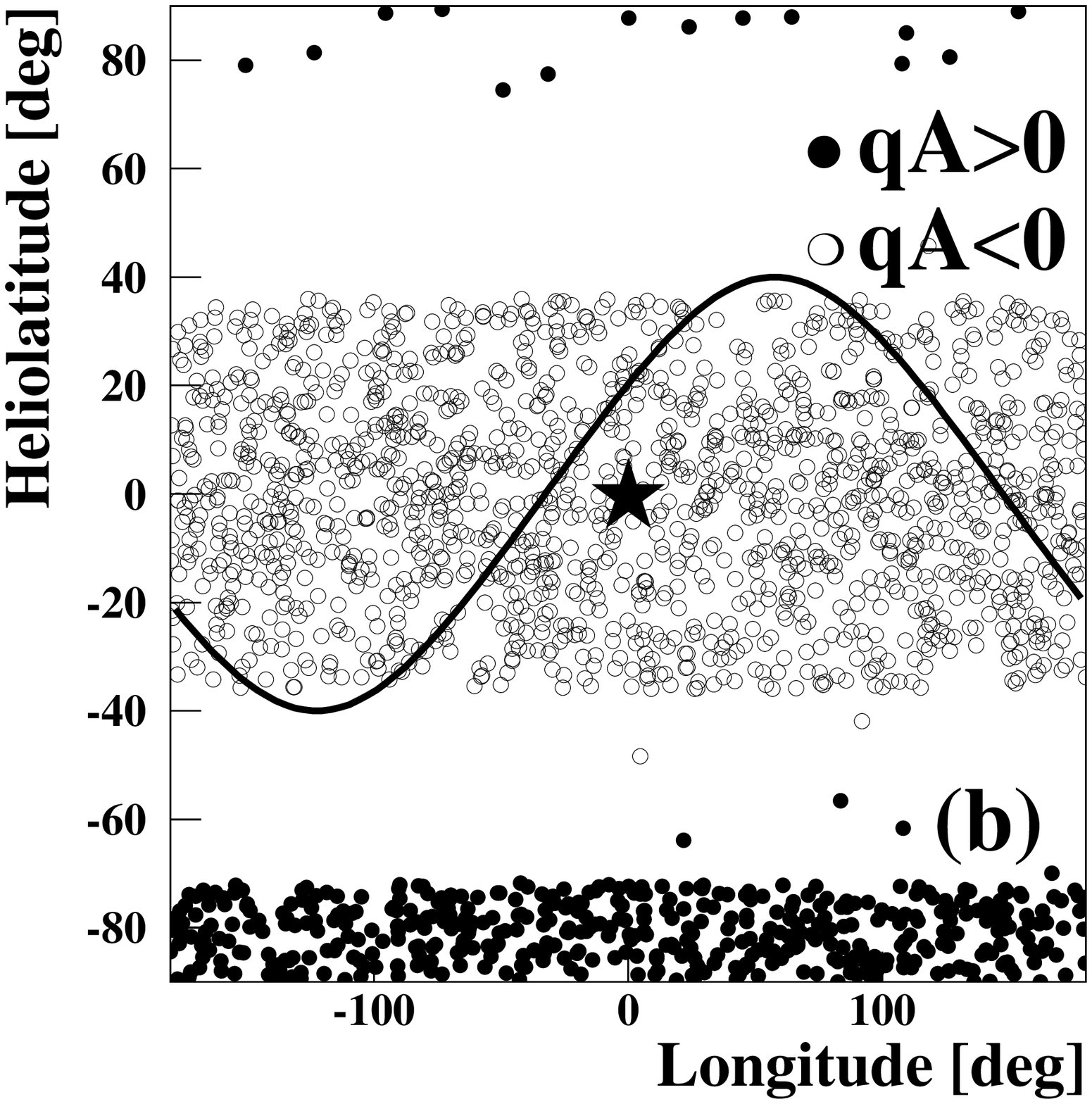}
  \end{center}
  \end{minipage}
  \caption{(a) Heliolatitude-longitude distribution by simulation for
 protons with energy 1 GeV at 1 AU when they exit the heliospheric
 boundary, 80 AU. The observer at 1 AU on the equatorial plane is
 located at a point northern the HCS; (b) Same as (a), but the observer
 is located at a point southern the HCS. The tilt angle near the sun is
 $40^\mathrm{o}$, and the solar activity heading for the maximum.}
 \label{distribution}
\end{figure}

Figure \ref{distribution} shows the heliolatitude-longitude distribution
of simulated particles when they exit the heliospheric boundary, where
the longitude is shown in the corotating coordinate system. The thick
line indicates the HCS at 1 AU when the particles start off from a point
(indicated by ``star'') at 1 AU on the equatorial plane. Latitudes of
the arrival points for $qA<0$ distribute around the HCS as expected,
because particles arrive at 1 AU along the HCS. The thickness of the
distribution of latitude ($\pm 30^\mathrm{o}$) reflects the structure of
spatiotemporal variation of the HCS. The structure of the HCS seen by
particles is different from particle to particle depending on the
arrival time. On the contrary, the latitude for $qA>0$ distribute around
the polar region, either northern (a) or southern (b) depending on
whether the observer is located at a point northern (a) the HCS or
southern (b), respectively. This tendency is also as expected, because
it is difficult for positive particles to penetrate through the HCS
during the journey from the outer heliosphere. The mean arrival time of
1 GeV protons is 4.5 days and 80 days for $qA>0$ and $qA<0$,
respectively. The short time of only 4.5 days for $qA>0$ suggests that
particles are less modulated. This relatively little modulation effect
is reflected in the fail to reproduce the observed spectrum for $qA>0$
as we mention later.

\begin{figure}[t]
  \begin{minipage}{0.5\hsize}
  \begin{center}
 \includegraphics[bb = 0mm 55mm 200mm 230mm, scale = 0.36]{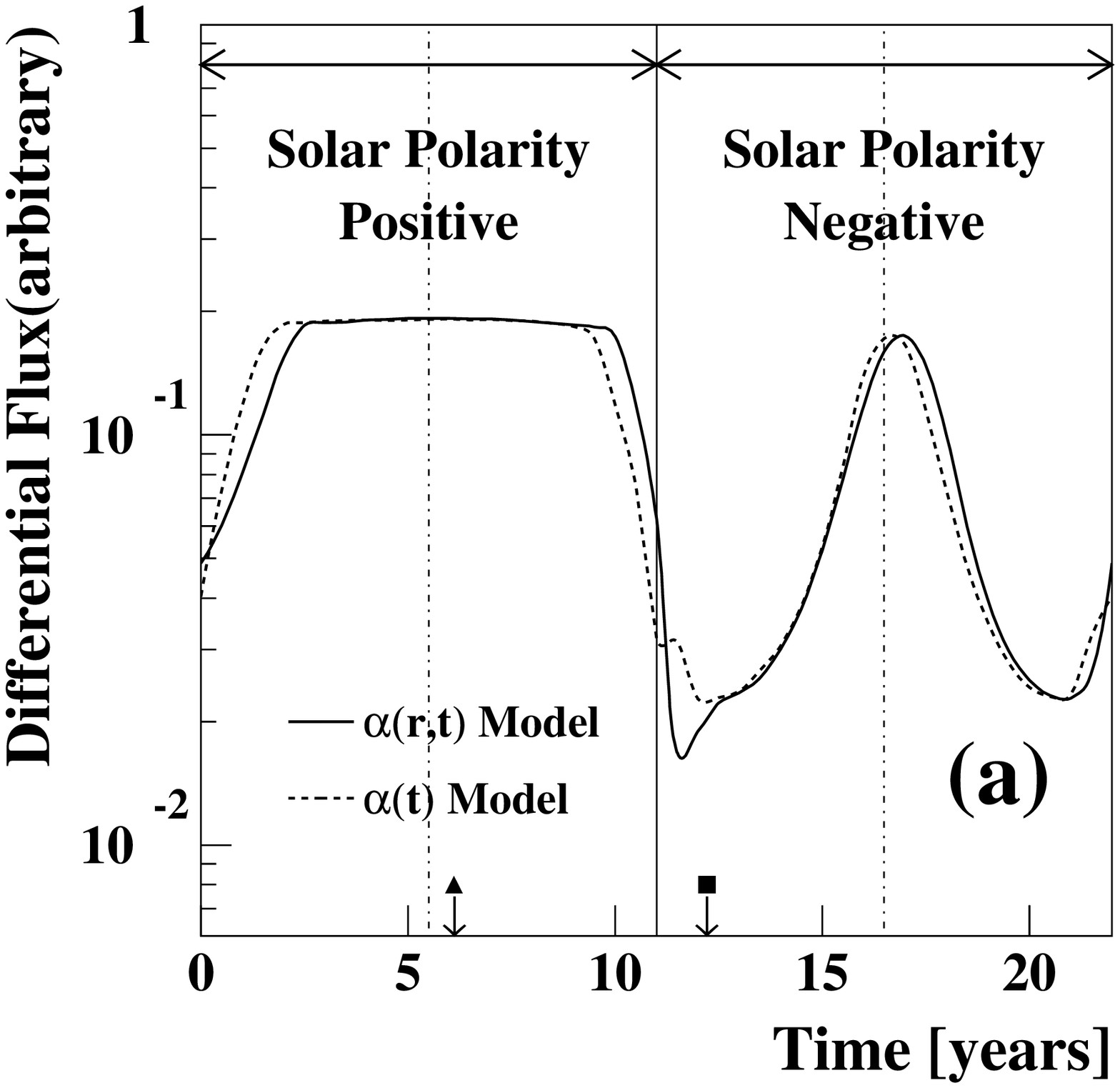}
  \end{center}
  \end{minipage}
  \begin{minipage}{0.5\hsize}
  \begin{center}
 \includegraphics[bb = 0mm 55mm 200mm 230mm, scale = 0.36]{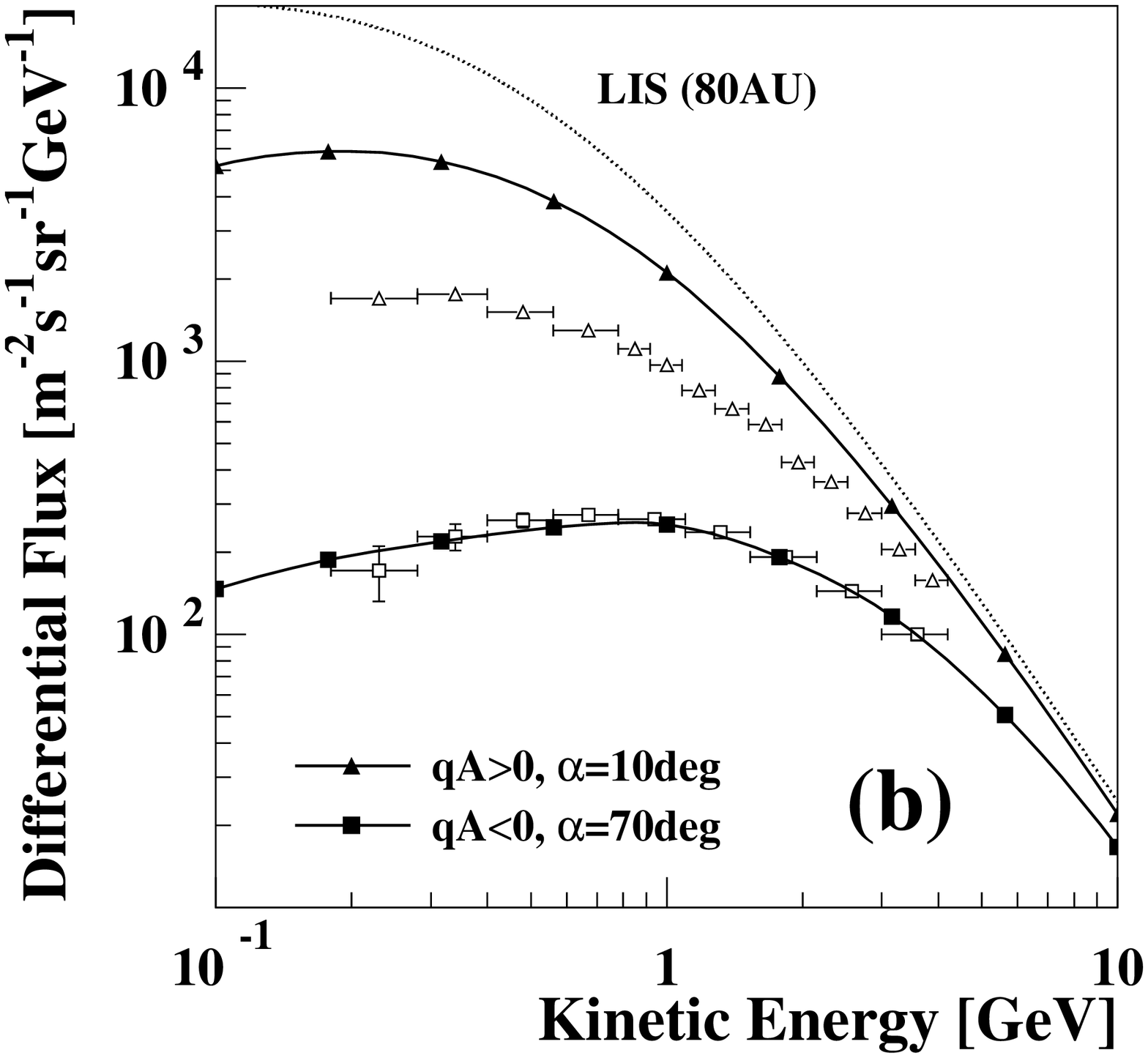}
  \end{center}
  \end{minipage}
  \caption{(a) Simulated variation of the 1 GeV proton intensities over the
 22-year modulation cycle; (b) Simulated proton energy spectra and
 observed proton energy spectra in 1998 and 2000.}
 \label{p22-spec}
\end{figure}

Figure \ref{p22-spec}a shows simulated variation of the proton
intensities over the 22-year modulation cycle. The solid line indicates
the result for the model of the HCS by eq.\ref{theta} ($\alpha (r,t)$
model). The dashed line shows the result for the HCS model ($\alpha (t)$
model) where the tilt angle varies with time but does not depend on
radius from the sun. In our $\alpha (r,t)$ model, the variation of the
HMF near the sun propagates with the solar wind velocity. It takes about
1 year for the signal of polarity change and change in the tilt angle
reach the heliospheric boundary. In $\alpha (t)$ model, on the contrary,
these changes reach instantaneously the boundary. This difference
between the two models in signal propagation is reflected in a time lag
of the proton intensity changes by $\alpha (r,t)$ model as seen in Figure
\ref{p22-spec}a. Our model succeeded in reproducing a 22-year modulation
cycle and in reproducing, though qualitatively, rather flat maximum for
$qA>0$ and a sharp maximum for $qA<0$ \cite{NM}.

Figure \ref{p22-spec}b shows proton energy spectra by our simulation (solid
line) and the spectra observed by BESS experiment (open symbol). The open
triangles refer to proton fluxes observed in 1998 \cite{Data1998} when
the tilt angle near the sun is about $10^\mathrm{o}$ and the solar
polarity is positive. The open squares refer to proton fluxes observed
in 2000 \cite{Data2000} when the tilt angle near the sun is about
$70^\mathrm{o}$ and the solar polarity is negative. The symbols for
simulated energy spectra refer to the phases marked by the same symbol
in Figure \ref{p22-spec}a. The view of the HCS corresponding to these
phases are shown in Figure \ref{cs}a and \ref{cs}c. The local
interstellar spectrum (LIS) of proton is assumed as $j_T \propto \beta
(T + 0.5mc^2)^{-2.6}$, where $T$ is kinetic energy and $m$ is the mass
of proton. The simulated energy spectra are normalized at 3.6 GeV to the
observation in 2000 \cite{Data2000}. Our model spectrum for $qA<0$
agrees quite well with BESS result, however our model fails to reproduce
the observed spectrum for $qA>0$. The failure of our model for $qA>0$
may come from inadequacies of the heliospheric structure in our
model. As we mentioned before, positive particles drift from polar
region within short times modulated rather weakly. If we take into
account the existence of the random transverse component of the HMF in
the polar regions \cite{Pole}, we may be able to reproduce the observed
spectrum because it is expected that the random transverse component
suppresses drifts in the polar regions. The actual variation of the HCS
is much irregular in contrast to our model. We have to consider the
actual dynamically changing HCS.

\section{Conclusions}

We have developed a time-dependent and three-dimensional code based on
the SDE method. The sample trajectories and distributions of arrival
points at the heliospheric boundary agree with the physical
expectation. Our simulation reproduced a 22-year modulation cycle which
is qualitatively consistent with observations. Our model spectrum for
$qA<0$ agrees with BESS result, however our model fails to reproduce the
observed spectrum for $qA>0$.

\section{Acknowledgements}

We would like to thank Prof. J. Kota and Prof. P. Evenson for their
comments and suggestions on this work.


\begin{thebibliography}{99}

\bibitem{Yamada}
Yamada Y., S. Yanagita, and T. Yoshida, Geophys. Res. Lett., 25,
	2353-2356, 1998.

\bibitem{Zhang}
Zhang M., Astrophys. J., 513, 409-420, 1999.

\bibitem{Drift}
Burger R. A. and M. S. Potgieter, Astrophys. J., 339, 501-511, 1989.

\bibitem{Sheet}
Jokipii J. R. and B. Thomas, Astrophys. J., 243, 1115-1122, 1981.

\bibitem{Kota}
Kota J. and J. R. Jokipii, 28th ICRC (Tsukuba), 3791-3794, 2003.

\bibitem{NM}
Webber W. R. and J. A. Lockwood, J. Geophys. Res., 93, 8735-8740, 1988.

\bibitem{Data1998}
Maeno T. et al., Astropart. Phys., 16, 121-128, 2001.

\bibitem{Data2000}
Asaoka Y. et al., Phys. Rev. Lett., 88, 051101, 2002.

\bibitem{Pole}
Jokipii J. R. and J. Kota, Geophys. Res. Lett., 16, 1-4, 1989.

\end{thebibliography}
\end{document}